\documentclass[10pt]{article}
\usepackage{graphicx}

\usepackage{wrapfig}

\usepackage{a4wide}  
\usepackage{url}  
\usepackage[utf8]{inputenc}  
\usepackage{latexsym}  
\usepackage{amsmath} 
\usepackage{amssymb} 
\usepackage{setspace} 
\usepackage[]{algorithm2e}
\usepackage{subfig}

\newtheorem{property}{Property}

\newcommand{\ov}{\textrm{ov}}

\newcommand{\ovv}{\overline{v}}

\newenvironment{proof}{\noindent {\it Proof.}}{$\Box$\vskip1ex}

\graphicspath{{figures/}}

\makeatletter
\renewcommand{\algocf@caption@boxruled}{%
  \hrule
  \hbox to \hsize{%
    \vrule\hskip-0.4pt
    \vbox{   
       \vskip\interspacetitleboxruled%
       \unhbox\algocf@capbox\hfill
       \vskip\interspacetitleboxruled
       }%
     \hskip-0.4pt\vrule%
   }\nointerlineskip%
}%
\makeatother

\begin{document}

\title{On improving the approximation ratio of the $r$-shortest common superstring problem}

\author{Tristan Braquelaire \thanks{LaBRI and CBiB, University of
    Bordeaux, France.} \and Marie Gasparoux \thanks{DIRO, Univ. of Montréal, Canada and LaBRI, University of Bordeaux,
    France.} \and Mathieu Raffinot \thanks{CNRS, LaBRI, University of Bordeaux,
    France.}\and Raluca Uricaru$\,^*$}

\maketitle


\begin{abstract}

The Shortest Common Superstring problem (SCS) consists, for a set of
strings $S = \{s_1,\cdots,s_n\}$, in finding a minimum length string that
contains all $s_{i, 1\leq i \leq n}$, as substrings. While a
$2\frac{11}{30}$ approximation ratio algorithm has recently been
published, the general objective is now to break the conceptual lower
bound barrier of $2$. This paper is a step ahead in this
direction. Here we focus on a particular instance of the SCS problem,
meaning the $r$-SCS problem, which requires all input strings to be of
the same length, $r$. Golonev {\em et al.} proved an approximation
ratio which is better than the general one for $r\leq 6$. Here we
extend their approach and improve their approximation ratio, which is
now better than the general one for $r\leq 7$, and less than or equal to $2$ up to
$r = 6$.

\end{abstract}  


\section{Introduction}

The Shortest Common Superstring problem (SCS) consists, for a set of
strings $S = \{s_1,\cdots,s_n\}$ over a finite alphabet $\Sigma$ (with
no $s_i$ substring of $s_j$), in constructing a string $s$ such that
any element of $S$ is a substring of $s$ and $s$ is of minimal length.
For an arbitrary number of strings $n$, the problem is known to be
NP-Complete~\cite{GALLANT198050, Garey1990} and
APX-hard~\cite{Blum:1994}. Lower bounds for the achievable
approximation ratios on a binary alphabet have been given
\cite{Ott1999,KarpinskiS13}, and the best approximation ratio so far
for the general case is $2 \frac{11}{30} \approx 2.3667$
\cite{Paluch14}, reached after a long series of improvements
\cite{l-tdstls-90, Blum:1994,KPS94,
  Armen1995,Armen199829,Breslauer1997340,
  Czumaj199774,Sweedyk:1999,TengY97, KaplanS05, PaluchEZ12, Mucha13}
leading to increasingly complex algorithms. 

An SCS greedy algorithm is known to reach good performances in
practice but its guaranteed approximation ratio has only been proved
to be $3.5$ \cite{KaplanS05}, while conjectured $2$.

The general objective in this algorithmic domain is now to break the
conceptual lower bound barrier of $2$.  This paper is a step ahead in
this direction. Here, we focus on a particular instance of the SCS problem,
$r$-SCS, when all strings of the set of strings are of length
$r$. Improving the SCS, and in particular the $r$-SCS approximation
ratio is interesting from both theoretical and practical reasons, in
view of their numerous applications, like in bioinformatics and more
precisely in the reconstruction of DNA sequences from {\em reads},
which are usually short, fixed-length DNA strings. Gallant {\em et
  al.}~\cite{GALLANT198050} showed that the $r$-SCS problem stays
NP-hard, except for the $2$-SCS case that can be solved in polynomial
time~\cite{CrochemoreCIKRRW10}. Golovnev {\em et
  al.}~\cite{GolovnevKM13} proposed an approximation ratio for the
$r$-SCS problem, which was better than the best general approximation
ratio ({\em i.e.,} the one for the general SCS problem) at the time
their article was published ($2 \frac{11}{23}$~\cite{Mucha13}) for $r
< 8$. However, in the meantime, the general approximation ratio has
been improved from $2 \frac{11}{23}$ to $2 \frac{11}{30}$, thus
canceling their result for $r = 7$. In this new context, the
approximation ratio proved by Golovnev {\em et al.} for the $r$-SCS
problem remains better than the actual general one for $r < 7.$

In this article, we  {\em extend} the approach of
\cite{GolovnevKM13} and exhibit a new approximation ratio for the $r$-SCS problem
$\beta(r)$, 
$$\beta(r) = \max_{0 \leq x \leq r-1} \left \lbrace \min \left \lbrace
\frac{4 + (r - 2)(r - x - 1)}{2(r -
  x)},\frac{(r^2-2r+2)-(r-1)x}{r-x},\frac{r-\frac{2}{3}x}{r-x} \right
\rbrace \right \rbrace \;$$ that allows us to remain competitive for
$r \leq 7$ and less than or equal to $2$ for $r \leq 6$

Note that some theoretical variations of SCS have also been studied
\cite{Yu16a}. Here we neither dwell on these studies since their focus
is far from ours, nor detail the greedy algorithm approximation
conjecture, which is a subject in itself
\cite{TARHIO1988131,KaplanS05,FiciKRRW16}.

\section{State of the art for the SCS and $r$-SCS problems}

Given the immense number of results that have been published on the
SCS problem, in this section we will only retrace those results that
are relevant to this work.

\begin{paragraph}{Preliminary notations and definitions} 

Let $\Sigma$ be a finite set of characters (or letters), $\varepsilon$
the empty word, and $w =w_1 \ldots w_p \in\Sigma^*$ a string and
$||w||=p$ its length, with $||\varepsilon||=0$. We denote
$\mbox{pref}(w,k), 1 \leq k \leq ||w||$ (resp. $\mbox{suff}(w,k)$) the
prefix (resp. suffix) of length $k$ of $w$ as the string $w_1 \ldots
w_k$ (resp. $w_{||w||-k+1} \ldots w_{||w||}$). We extend our notations
with $\mbox{pref}(w,0)=\mbox{suff}(w,0)=\varepsilon$.

For two strings $u,v$ we define the \emph{maximum overlap} of $u$ and $v$,
denoted $\mbox{\em ov}(u,v)$, as the longest suffix of $u$ that is also a prefix
of $v$.

The \emph{overlap graph} built on a set of strings $S$ is a complete
directed graph with the vertex set $V=S$ and the edges set
$E=\{e_{i,j}=(s_i,s_j) | \forall s_i, s_j \in V\}$, with label
$l(e_{i,j})=\ov(s_i,s_j)$ and weight $w(e_{i,j})=||\ov(s_i,s_j)||.$
Figure \ref{fig:overlapgraph} shows the overlap graph built on $SE$,
where $SE=\{\mbox{ACGCA},\mbox{CGCAT},\mbox{GCATG},
\mbox{CGCAG},\mbox{CAGTC}, \mbox{CAGCA},\mbox{CATAA}\}$ is a set of
strings that will be of use for illustrative purposes throughout this
article. As in this work we focus on the $r$-SCS problem, note that,
without loss in generality, we take $SE$ as being a set of $5$-length
strings ($r=5$).

\begin{figure}[t]
  \centering
   \includegraphics[width=0.65\linewidth]{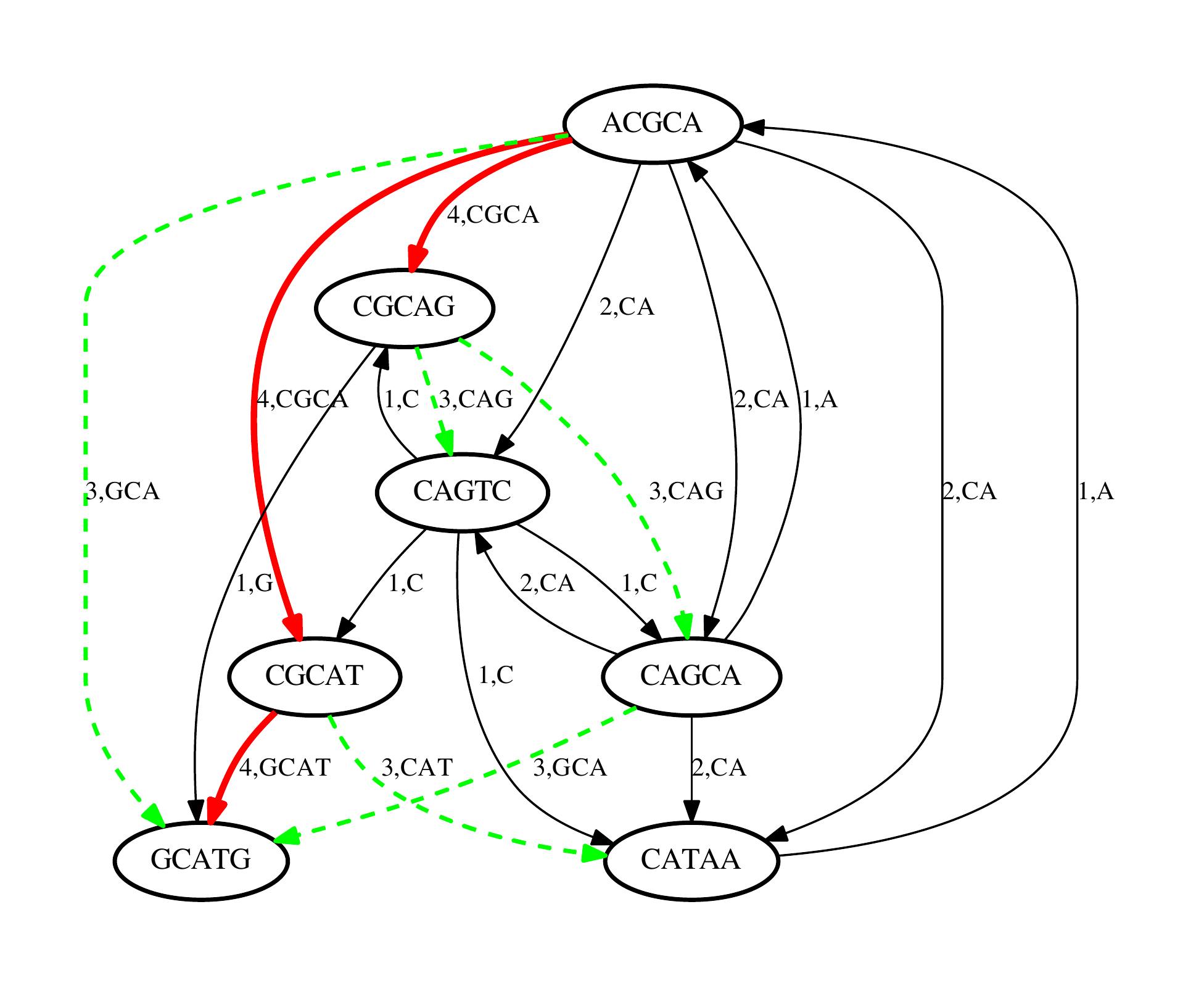}
  \caption{Example of an overlap graph on $SE$. Large arrows in red
    correspond to $(r-1)$-length overlaps, while middle width dashed green ones represent
    $(r-2)$-length overlaps. The graph is complete, but for the sake of clarity $0$-weighted edges are not represented in the figure.}
  \label{fig:overlapgraph}
\end{figure}
\end{paragraph}

\begin{paragraph}{Hamiltonian path approximation}

It is straightforward to see that finding a solution to the SCS
problem comes to computing a Hamiltonian path of maximum weight (named
$H$ below) in the overlap graph (described above). Indeed, $H$ would
directly lead to a shortest superstring solution for the SCS problem,
whose compression value (with respect to the naive superstring obtained by simply concatenating the $n$ strings) would be equal to the weight of $H$, denoted by
$w(H)$. 

The best existing approximation algorithm \cite{Kaplan:2005}
for computing a weighted hamiltonian path (derived from the asymmetric
maximum traveling salesman path, MAX\_ATSP), gives a hamiltonian path
whose length is at least $\frac{2}{3}$ of the weight of the longest
path, \emph{i.e.,} $\frac{2}{3} w(H).$ Therefore, this gives a
superstring solution that is a $2.5$-approximation for the SCS
problem, which is far from the actual best known approximation ratio
($2 \frac{11}{30} \approx 2.3667$ \cite{Paluch14}) but it is important
for the remaining of this paper.
\end{paragraph}

\subsection{$r$-SCS problem : the approach of Golovnev {\em et al.}}

In this paper we {\em extend} the approach of Golovnev {\em et
  al.}~\cite{GolovnevKM13} for the $r$-SCS problem, which we summarize
in the remaining of this section. Golovnev {\em et al.} use a
$(r-1)$-spectrum in order to translate the initial instance of the
$r$-SCS problem in a $2$-SCS instance, which they exactly solve with
the approach described in \cite{CrochemoreCIKRRW10}; this gives them a
solution that, once translated back to the original problem,
represents a good approximation of the optimal superstring for the
original problem, given that the optimal is small.

In the remaining of this paper, we denote $\mbox{OPT}(S)$ for $S =
\{s_1,\cdots,s_n\}$ as the length of a shortest supertring of $S.$
Clearly, when considering the $r$-SCS problem, $\mbox{OPT}(S)=rn-w(H)$
(see the paragraph on the {\em Hamiltonian path approximation}).

\subsubsection{$k$-spectrum and de Bruijn graphs}
\label{kspectrum}

Given that a $k$-mer is a string of length $k$, in Golovnev {\em et
  al.}  the notion of $k$-spectrum of the input set is defined as the
set of $k$-mers issued from the sequences of the input set. 

{\em De Bruijn graphs} are largely employed in next
generation sequencing (NGS) data analysis, and specifically in genome
assembly, as they display interesting properties like providing an
intrinsic succinct representation of the data, and enabling the
implementation of efficient methods for computing a superstring
solution (which represents a reasonable approximation of the original
genome sequence). A de Bruijn graph modeling a set of strings $S$ is built on the $k$-spectrum of $S$ as following: nodes are $k$-mers and oriented edges connect two $k$-mers if they overlap on exactly $k-1$ characters.

\begin{figure}[h!]
  \centering
\includegraphics[width=6.5cm]{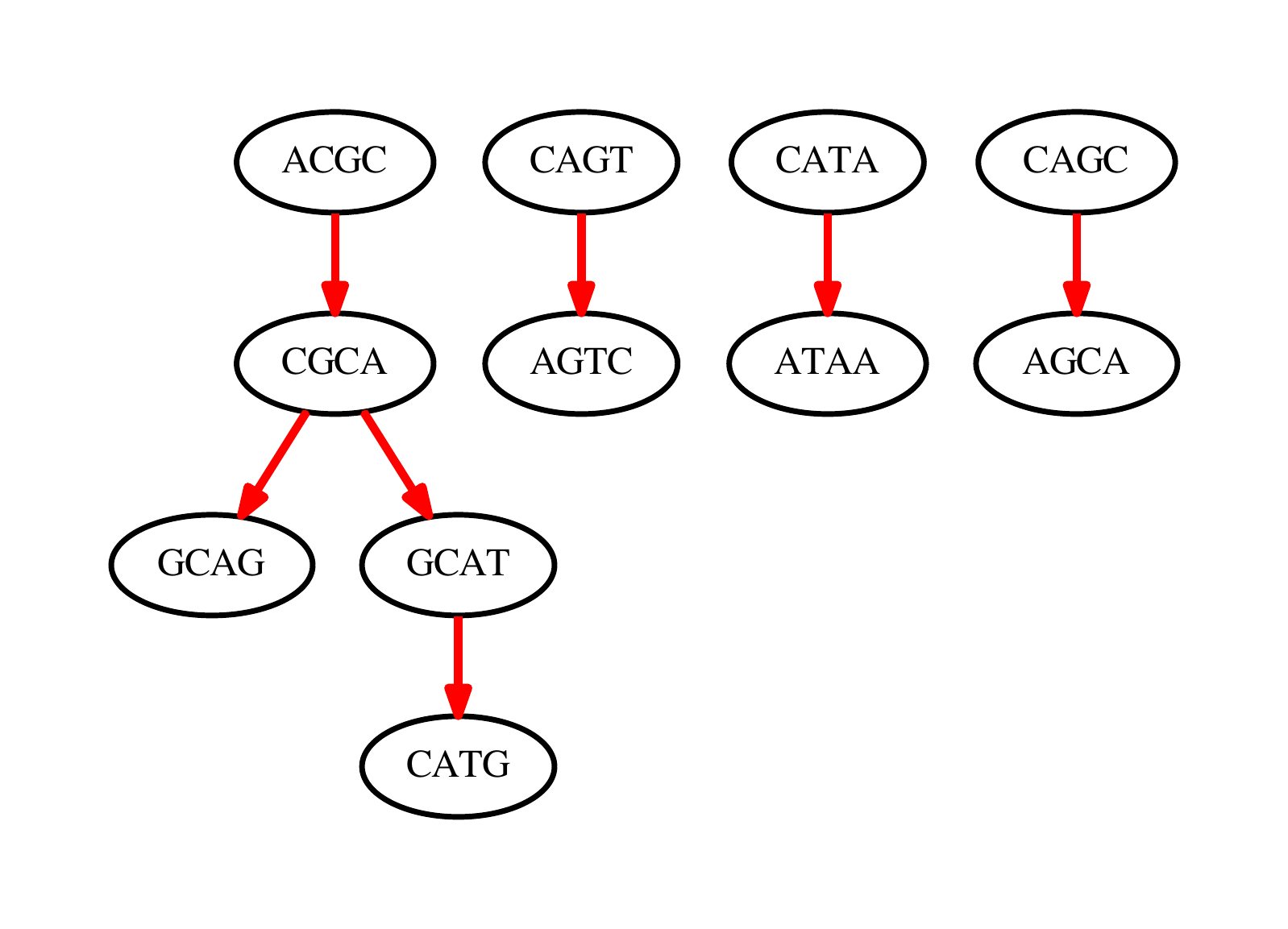} \quad \includegraphics[width=7cm]{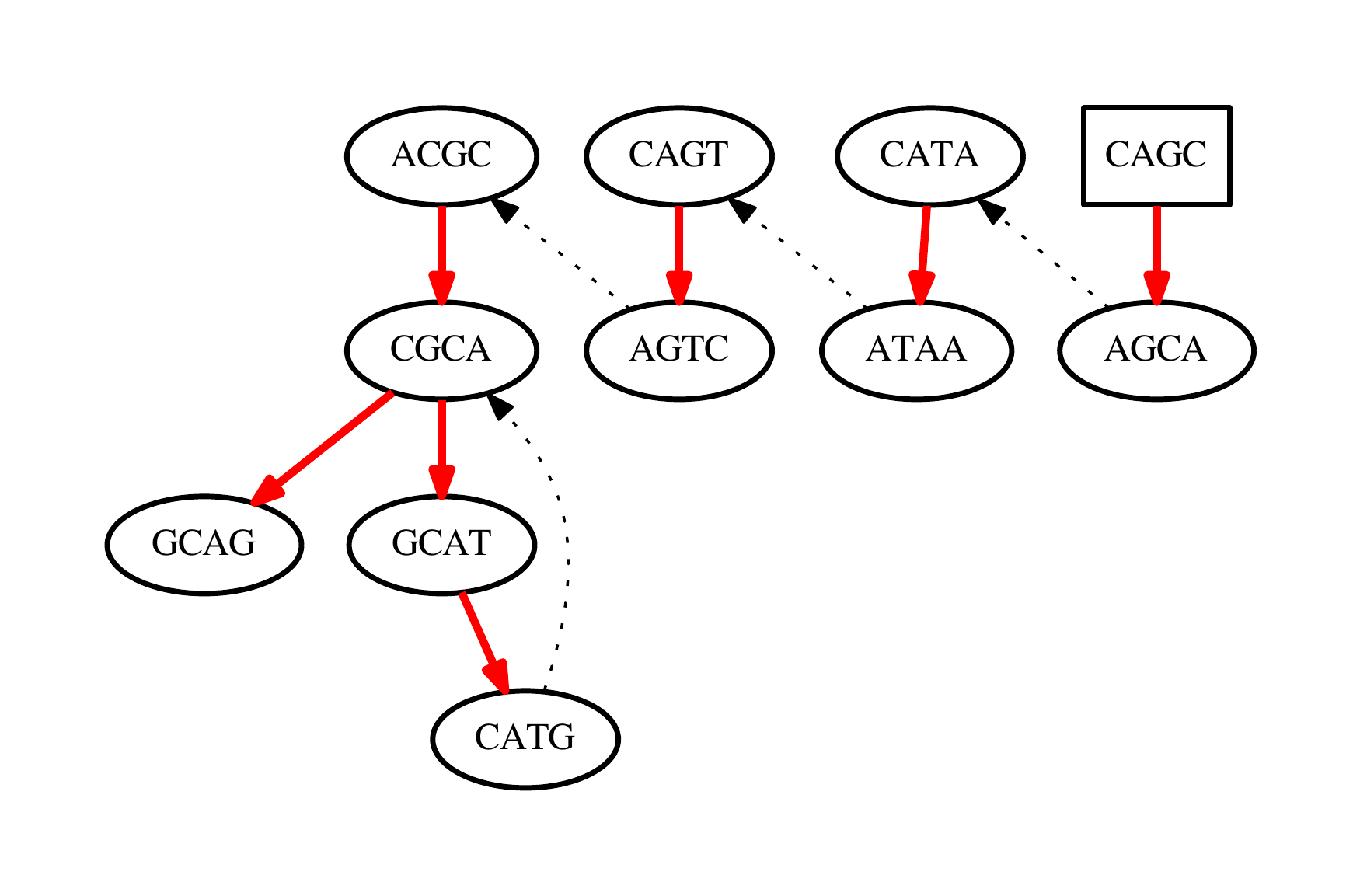}
  \caption{(left) A de Bruijn graph (edges correspond to strings in the initial set) built on the $4$-spectrum of the set $SE=\{\mbox{ACGCA},\mbox{CGCAT},\mbox{GCATG}, \mbox{CGCAG},\mbox{CAGTC}, \mbox{CAGCA},\mbox{CATAA}\}$. (right) An example of an eulerian path (with minimal additional edges depicted by dashed lines) on this graph (right figure). The boxed node is the first node of the path and traversing the path leads to the
    following superstring:
    $\tau_{SE}=\mbox{CAGCACATAACAGTCACGCATGCGCAG}.$}
  \label{fig:deBruijnEulerian}
\end{figure}

In this work, as in \cite{GolovnevKM13}, the de Bruijn graph is used in a particular context : given an initial set of strings of length $r$, a de Bruijn graph is built on the $(r-1)$-spectrum corresponding to this set of strings. Figure~\ref{fig:deBruijnEulerian} (left) shows this kind of de Bruijn graph built on the $4$-spectrum of the set $SE$ (containing strings of length $5$).

\subsubsection{$2$-SCS problem}
\label{sec:2scsexplanations}

The $2$-SCS problem is a particular case of the $r$-SCS problem when
$r=2$, which deserves special attention since it has been shown in
Crochemore {\em et al.}~\cite{CrochemoreCIKRRW10} that it is solvable
in polynomial time, even when considering multiplicities (meaning that
the strings must appear in the resulting superstring a given number of
times). As our approach capitalizes on some technical parts from the
method presented in \cite{CrochemoreCIKRRW10}, we give some insights
on this method in the following of this section.

Let us consider the set of strings $S_2$ (composed of strings of
length $2$), where $S_2 = \{s_1 = \alpha_1\beta_1, s_2 =
\alpha_2\beta_2, \ldots, s_n = \alpha_n\beta_n\},$ with $\alpha_i$ and
$\beta_i$ two characters. Let us also
take $m_i$, positive integers indicating the {\em multiplicity} of the
corresponding string $s_i$.

\begin{figure}[h!]
  \centering
   \includegraphics[width=0.40\linewidth]{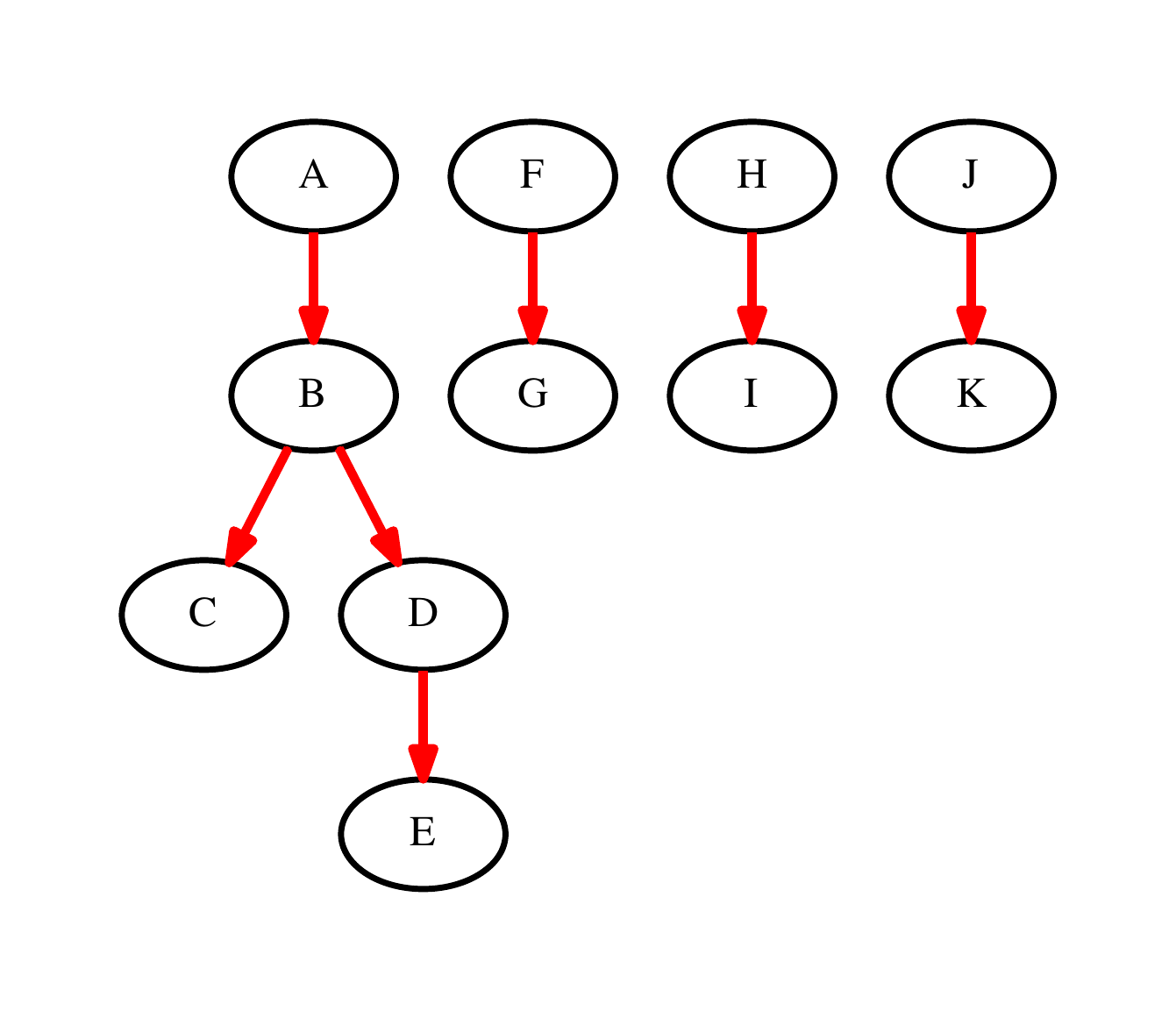} \quad \includegraphics[width=0.43\linewidth]{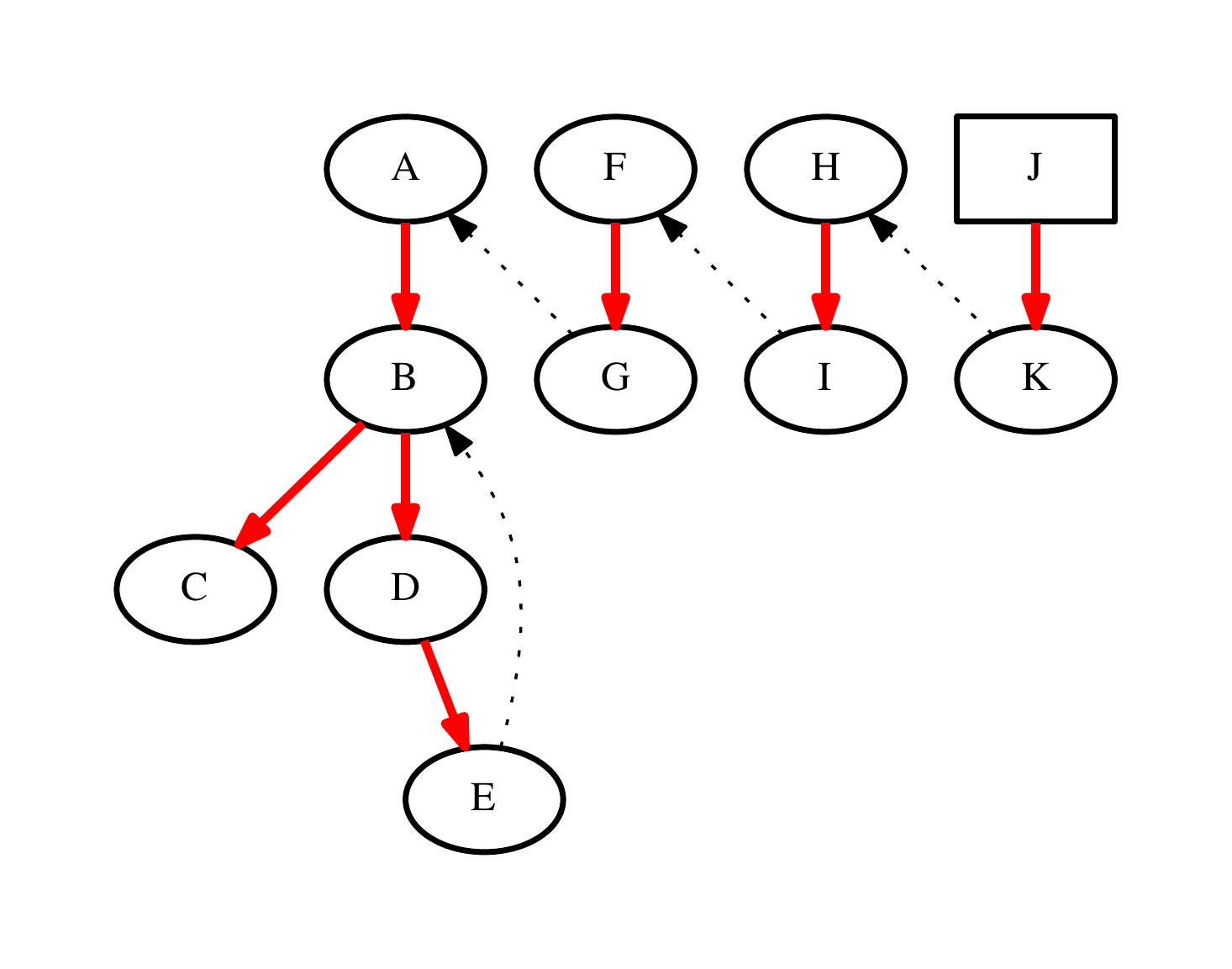}
  \caption{Illustration of the resolution of the 2-SCS problem on the set $\{
    \mbox{AB},$ $\mbox{BC},$ $\mbox{BD},$ $\mbox{DE},$ $\mbox{FG},$
    $\mbox{HI},$ $\mbox{JK} \}$, by building an eulerian path (with minimal additional edges). The resulting superstring can be obtained by traversing the eulerian path and concatenating the labels :
    J$\rightarrow$K$\rightarrow$H$\rightarrow$I$\rightarrow$F$\rightarrow$G$\rightarrow$A$ \rightarrow$B$\rightarrow$D$\rightarrow$E$\rightarrow$B$\rightarrow$C
    $=$ JKHIFGABDEBC.}
  \label{fig:2scs}
\end{figure}

In order to compute a superstring solution on this set, the approach
of \cite{CrochemoreCIKRRW10} is to build an oriented (possibly)
multi-graph $G$ from $S_2$ where nodes correspond to $\{\alpha_i |
1\leq i \leq n\}\cup \{\beta_i |1\leq i \leq n\}$ and each string $s_i
= \alpha_i\beta_i$ gives $m_i$ oriented edges from $\alpha_i$ to
$\beta_i$. 

Then, the graph $G$ is completed with a
minimum number of oriented edges to form an eulerian path
$EP$, that is a path passing through each and every edge exactly
once. A superstring is then obtained by following the path $EP$ and by
concatenating the characters labeling each node the path goes
through. Figure~\ref{fig:2scs} illustrates the way this approach works
on the set $\{ \mbox{AB},$ $\mbox{BC},$ $\mbox{BD},$ $\mbox{DE},$
$\mbox{FG},$ $\mbox{HI},$ $\mbox{JK} \}.$

\subsubsection{From $r$-SCS to $2$-SCS and back}
Golovnev {\em et al.} translate an $r$-SCS instance on a set $S$
(composed of strings of length $r$) in a $2$-SCS instance, by computing
the $(r-1)$-spectrum of $S$ and building a de Bruijn graph on this set of $(r-1)$-mers. By assigning a character to each $(r-1)$-mer, a string in $S$ originally of length
$r$, becomes a string of length $2$ in the novel alphabet.
Next, they exactly solve the $2$-SCS problem with the method described
in section~\ref{sec:2scsexplanations}, based on the graph
illustrated in Figure~\ref{fig:2scs}, obtained from the de Bruijn graph on the $(r-1)$-spectrum of $S$. They eventually expand the
resulting sequence (built on the new alphabet) by replacing each two
letters (\emph{i.e.,} two original $(r-1)$-mers) connected by an edge in the graph with their corresponding $r$-mer. This
leads to a superstring solution for the original $r$-SCS problem, not necessarily optimal, named
$\tau$; see Figure~\ref{fig:deBruijnEulerian} (right) for an illustration of the $r$-SCS solution computed with this procedure.

The length of $\tau$ can be compared to $OPT$ (the length of the optimal superstring solution for the $r$-SCS problem), by observing that the number of $(r-1)$-overlap edges in an optimal weighted hamiltonian path ($H$) on the overlap graph, is necessarily less than or equal to the number of edges of the corresponding de Bruijn graph built on the $(r-1)$-spectrum. 

Indeed, it is trivial to see that the number of edges in the de Bruijn graph is equal to the number of strings in $S$.  See \cite{GolovnevKM13} for more details on this.

\section{Two steps hierarchical SCS approximation}
\label{gen2scs}

In this section we tackle the $r$-SCS problem for a set of strings $S$ with a hierarchical $2$-step procedure that generalizes the $2$-SCS approach introduced by Golovnev {\em et al}.

\subsection{Overview of our method}

\begin{figure}[ht]
  \centering
  \includegraphics[width=7.5cm]{eulerian1.pdf} \quad
 \includegraphics[width=6.5cm]{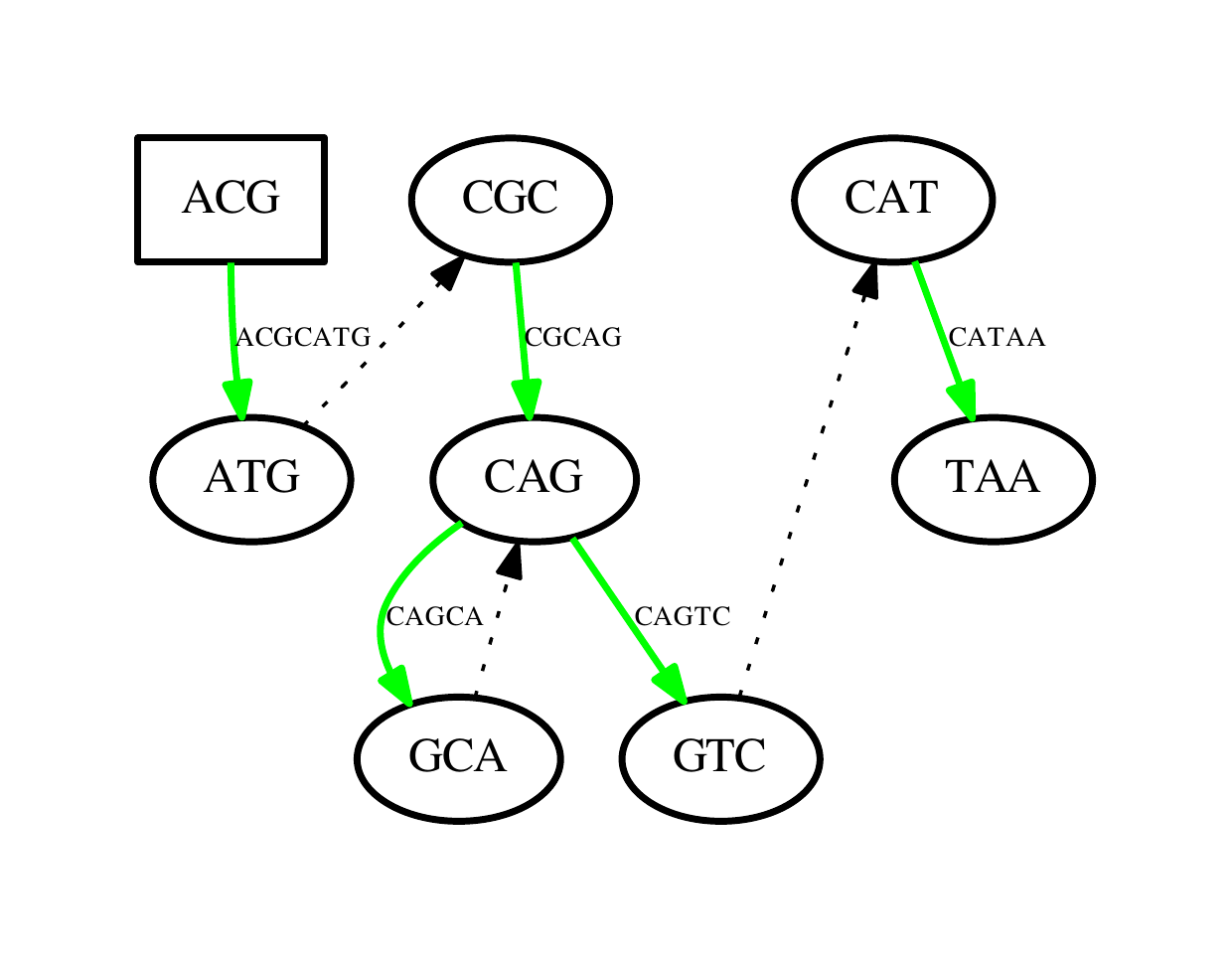}
  \caption{(left) Example of an eulerian path (depicted by dashed lines) with minimal additional edges,
    built on the $(r-1)$-spectrum of $SE.$ The boxed node is the
    start of the path. The expression of the path leads to the
    following set $SE'$ of contigs: $\mbox{CAGCA},$ $\mbox{CATAA},$
    $\mbox{CAGTC},$ $\mbox{ACGCATG},$ $\mbox{CGCAG}.$ (right) A
    minimal additional edges eulerian path built on the
    $(r-2)$-spectrum on the set of contigs $SE'.$ This second eulerian path starts with the boxed node $\mbox{ACG}$ and ends with $\mbox{TAA},$ thus producing the following superstring solution on $SE$ : \mbox{ACGCATGCGCAGCACAGTCCATAA}.} 
  \label{fig:eulerian}
\end{figure}

In \textbf{the first step} of our approach we apply the same translation as in \cite{GolovnevKM13}, meaning that from the original $r$-SCS problem, by using the $(r-1)$-spectrum of $S$, we obtain a $2$-SCS instance. After computing an optimal solution for the $2$-SCS problem with the algorithm presented in \cite{CrochemoreCIKRRW10}, and then applying the reverse translation, we obtain a first, unsophisticated solution. This solution is the same as the one output by \cite{GolovnevKM13}, but their method stops here.  In our case, we continue the initiated process by subsequently generating \emph{a set of ``contigs''}, {\em i.e.,} substrings of the superstring solution obtained by embedding strings from $S$ if overlapping on exactly $r-1$ characters. Note that contigs correspond to paths in the de Bruijn graph, and that the length of a contig is at least $r$. This set of contigs is computed from the initial superstring solution by cutting the superstring up in chunks, every time the connection is not due to an edge in the graph but rather to an edge added by the eulerian path resolution procedure from \cite{CrochemoreCIKRRW10} (represented by dashed lines in Figure \ref{fig:eulerian}~(left)). Figure \ref{fig:eulerian}~(left) illustrates this process on our example set $SE=\{\mbox{ACGCA},\mbox{CGCAT},\mbox{GCATG}, \mbox{CGCAG},\mbox{CAGTC}, \mbox{CAGCA},\mbox{CATAA}\}$ resulting in the set of contigs $SE' =\{$\mbox{ACGCATG}, \mbox{CGCAG}, \mbox{CAGTC}, \mbox{CAGCA}, \mbox{CATAA}$\}.$

\paragraph{Second step on the generalized spectrum of contigs} 

For the next step we extend the notion of $k$-spectrum to take as input a set of contigs
of possibly different lengths, but all greater than $k+1$ : the $k$-mers composing this new type of $k$-spectrum are the prefixes and suffixes of size $k$ of the input contig sequences. In our case, we compute a $(r-2)$-spectrum on the set of contigs issued from the first step of the method.

We then build a kind of de Bruijn graph for which the nodes come from the $(r-2)$-spectrum of the contigs and for each  contig sequence $w$, we add an oriented
edge from $\mbox{pref}(w,(r-2))$ to $\mbox{suff}(w,(r-2))$ labeled by
$w$. Figure \ref{fig:eulerian}~(right) shows such a graph built on the 
$(r-2)$-spectrum of the set of contigs $SE'.$ Finally, as in the first step, an eulerian path with minimal additional edges is computed on this graph, which gives a novel superstring solution for the $r$-SCS problem, that we call $\gamma.$

\subsection{Algorithm}

The intuition behind our algorithm is to push further the approach of Golovnev {\em et al.}
by additionally taking into account the $(r-2)$-overlap edges from the overlap graph built on $S$. However, this extension is not straightforward since (a) choices of $(r-1)$-edge paths are made in the first step of our algorithm, which cannot be reconsidered in the following steps; these $(r-1)$-edge paths selected in the first step prevent us from using some $(r-2)$ edges from the overlap graph, typically those branching inside a contig,
and (b) the contigs possibly have different lengths (thus the translation into à $2$-SCS instance is not straightforward). The algorithm, which is described more formally below, is illustrated in Figure~\ref{fig:eulerian} on our example set $SE$. 

\vspace{0.5cm}

\RestyleAlgo{boxruled}
\LinesNumbered
\begin{algorithm}[H]
\setstretch{2}

\KwData{$S=\{s_1,s_2,\ldots ,s_n\}$ a set of $n$ strings of length $r$}
\KwResult{$\gamma$ a superstring of the strings in $S$}

Build the de Brujn graph $dB(r-1)$ on $S$ as following: nodes are $(r-1)$-mers of all $s_i, 1 \leq i \leq n$; for each $s_i, 1 \leq i
\leq n$ add an edge from the node $\mbox{pref}(s_i,(r-1))$ to
the node $\mbox{suff}(s_i,(r-1))$. The graph built in this manner can easily be transposed into a $2$-SCS instance.

Solve the $2$-SCS instance with the algorithm from Crochemore {\em et al.}, giving a minimum size eulerian path in $dB(r-1)$. Build the corresponding contigs, $c_1 \ldots c_k$ of
varying sizes (at least $r$), by removing all edges added by the eulerian computation procedure
on $dB(r-1)$. We denote by $S'=\{c_1, \ldots ,c_k\}$ this new set of strings.

Build the special type of de Bruijn graph (described above), $dB'(r-2)$, on the generalized $(r-2)$-spectrum of $S'$: all the $(r-2)$-mers at the extremities of $c_i, 1 \leq i \leq k$, as nodes in the graph, and edges (corresponding to the contigs $c_i$), connect  a node $\mbox{pref}(c_i,(r-2))$ to a node
$\mbox{suff}(c_i,(r-2))$.

Solve the new $2$-SCS instance on $dB'(r-2)$ with the algorithm by Crochemore {\em et al.}
and output the corresponding superstring named $\gamma.$
\caption{Computing $\gamma$ superstring solution}
\label{AlgorithmGAMMA}
\end{algorithm}
\vspace{0.5cm}

\subsection{Analysis}

Golonev {\em et al.} based their analysis on the property that the
eulerian path they build on the de Bruijn graph for producing their superstring solution $\tau$ (corresponding to items $1$ and $2$ in Algorithm~\ref{AlgorithmGAMMA}) contains all $(r-1)$-overlap edges from the overlap graph, and thus at least as many as the $(r-1)$ edges taken in the hamiltonian path $H$ built on the overlap graph.

Our 2 steps algorithm is more difficult to analyse given that
our $\gamma$ superstring solution mixes $(r-1)$ and $(r-2)$ overlaps, and, if
it also contains at least as many $(r-1)$ overlaps as $H$, the number of
$(r-2)$ overlaps can be less than that used in $H$, due to the fact
that we do not consider all $(r-2)$ overlaps when building our
generalized $(r-2)$-spectrum (see Section \ref{gen2scs}). The
following property allows us to compare the number of $(r-1)$ and
$(r-2)$ edges in $H$ and in the eulerian path producing $\gamma.$

We denote $v$ the number of edges of weight at most $(r - 3)$ in $H$
and $\ovv=n-1-v$ the number of edges with weights $(r-1)$ and $(r-2)$ in $H.$

\begin{property}
Let $t$ be the number of $(r-1)$ and $(r-2)$ edges contained in
$\gamma$. Then $t\geq \ovv.$
\label{prop:missingedge}
\end{property}

\begin{figure}[t]
  \centering
   \includegraphics[width=0.30\linewidth]{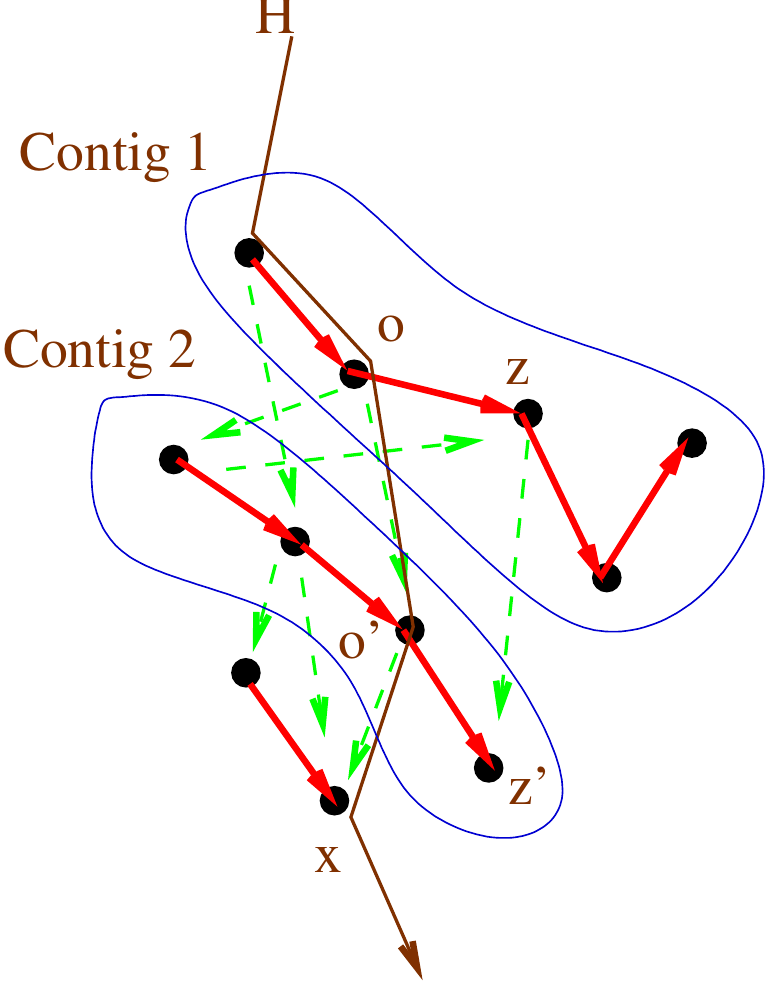} \quad \includegraphics[width=0.30\linewidth]{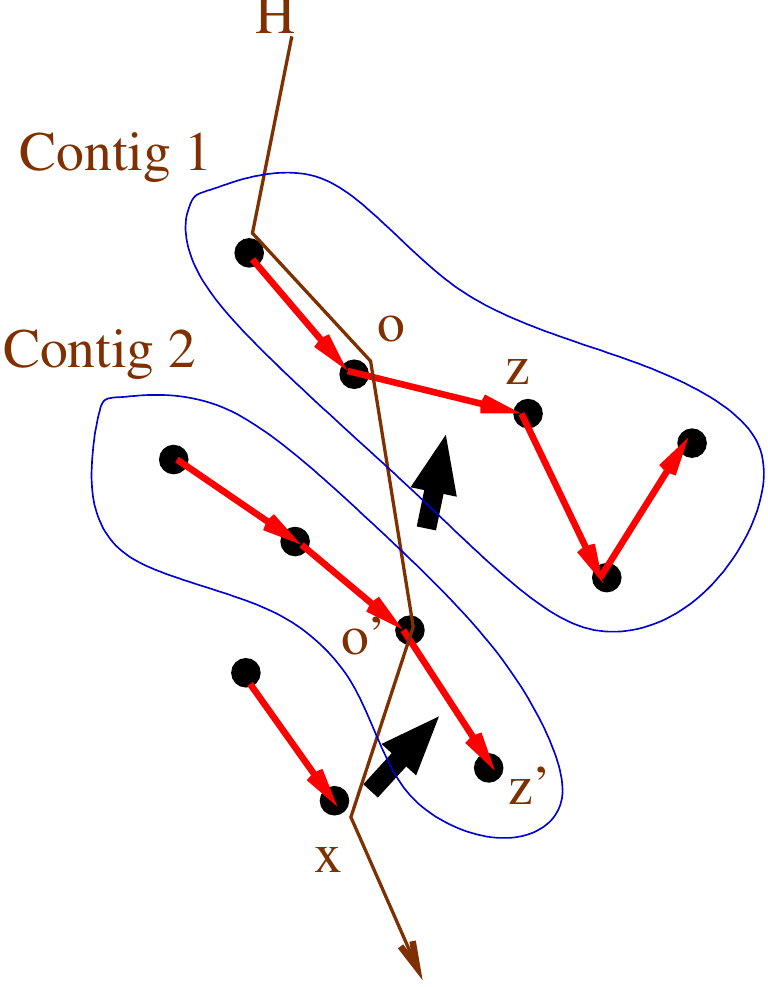}
  \caption{Illustration of the proof of Property~\ref{prop:missingedge}. The
    green dashed edges are  $(r-2)$-overlap edges that are not considered in the generalized
    $(r-2)$-spectrum. In the case where $H$ passes through such an edge
    $(o,o')$ (resp. $(o',x)$), we can uniquely associate to this edge
    the $(r-1)$-overlap edge $(o,z)$ (resp. $(o',z')$) represented
    by thick black edges.}
  \label{fig:missingedges}
\end{figure}

\begin{proof}
Let $t_1$ (resp. $t_2$) the number of $(r-1)$ (resp. $(r-2)$)
edges contained in $\gamma$ and symmetrically $\ovv_1$ (resp. $\ovv_2$)
the number of $(r-1)$ (resp. $(r-2)$) edges contained in $H.$ We
know that $t_1 \geq \ovv_1$ and let $\Delta_1=t_1-\ovv_1 \geq 0.$ It is straightforward that in the case where $t_2 > \ovv_2$  the property holds. Assume now
that $t_2 \leq \ovv_2$ and let $\Delta_2=\ovv_2-t_2 \leq 0$. We prove that in
this case $\Delta_1 \geq \Delta_2.$

Indeed, let us suppose that $H$ in the overlap graph passes through an $(r-2)$-overlap edge $l=(o,d)$ (where $o$ and $d$ are nodes in $OG$ corresponding to strings from $S$) that does
not appear in the de Bruijn graph on the generalized $(r-2)$-spectrum (built on the set of
contigs derived from the $(r-1)$-spectrum). This means that the origin
$o$ of $l$ lies inside a contig $c =s_u s_{u+1} \ldots s_v$ : there exists $u < j < v$ such that $o = s_j$ (beginning and end of $c$ non included); Figure \ref{fig:missingedges} illustrates this
case. Note that given the first step of our algorithm, if  $o = s_j$ precedes $s_{j+1}$ in a contig $c$, this corresponds to a $(r-1)$ edge from $o=s_j$ to $s_{j+1}$ in the $OG$. 
As $H$ is hamiltonian, $H$ only passes once through $o$ and thus (a)
cannot pass through the $(r-1)$ edge $(o,s_{j+1}); $ and (b) cannot pass twice through a $(r-2)$ edge from $o$. Thus, we can exactly 
map each $(r-2)$ edge from $H$, not appearing in the generalized
$(r-2)$-spectrum, to a $(r-1)$ edge of $\gamma$, not belonging
to $H,$ which proves that $\Delta_1 \geq \Delta_2$. Therefore $t=t_1+t_2 \geq \ovv=\ovv_1+\ovv_2.$\end{proof}

Based on Property~\ref{prop:missingedge} we are able to bound the length
of $\gamma$ relatively to $w(H)$.
Indeed, as we are not able to compare exactly the number of $(r-1)$
and $(r-2)$ edges between $\gamma$ and $H$, we consider these edges
as a whole, by counting an $(r-1)$-overlap as an $(r-2)$ one. Thus, compared to Golonev {\em at al.}, our approach further capitalizes on large overlap edges in $OG$  but introduces a
bounded approximation in the analysis.\\

\begin{figure}[h]
  \centering
  \includegraphics[width=15cm]{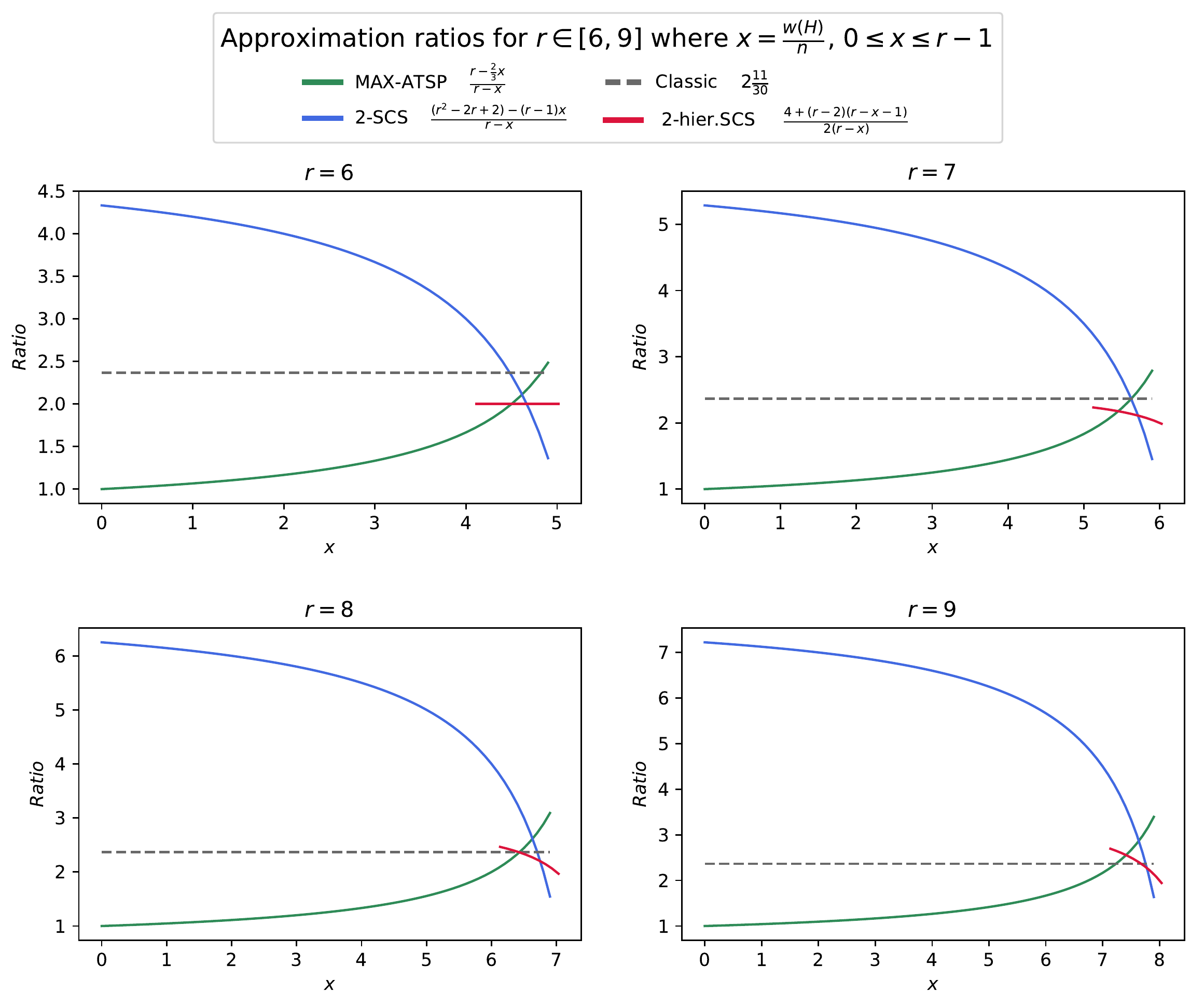}
  \caption{Comparison of our bound (in red) to MAX-ATSP, to the
    best general bound of $2\frac{11}{30}$, and to that of Golonev {et al.}}
  
  \label{fig:plots_r6_r9}
\end{figure}

We recall that the number of edges of weight either $(r - 1)$ or $(r -
2)$ in $H$ is $(n - 1 - v)$.  Then $w(H) \leq (r - 1)(n - 1 - v) + (r
- 3)v$ and hence
\[v  \leq \frac{(r - 1)(n - 1) - w(H)}{2} \;.\]
Since
\begin{itemize}
\item a shortest superstring for $S_{2SCS}$ ($2$-SCS instance obtained from the original $r$-SCS instance) contains $t\leq \overline{v}$ overlaps in $H$, and
\item the maximal length of a superstring for $S_{2SCS}$ is $2n$
\end{itemize}
then the length of a shortest superstring for $S_{2SCS}$ is at most
\[2n - (n - 1 - v)\;.\]
Hence, the corresponding superstring $\gamma$  for $S$ is of length at most
\[\max(rn - (r - 1)(n - 1 - v), rn - (r - 2)(n - 1  - v))\]
As $rn - (r - 2)(n - 1 - v) - (rn - (r - 1)(n - 1  - v) = n - 1 - v$ and $v < w(H) \leq n - 1$,
the superstring is of length at most
\begin{align*}
rn - (r - 2)(n - 1 - v) &\leq rn - (r - 2)(n - 1) + (r - 2)v \\
&\leq rn - (r - 2)n + \frac{r - 2}{2}((r - 1)(n - 1) + 2) - \frac{r - 2}{2}w(H) \\
&\leq 2n + \frac{r - 2}{2}((r - 1)(n - 1) + 2) - \frac{r - 2}{2}w(H) \\
&\leq \frac{4n + (r - 2)((r-1)n - w(H))}{2} \qquad \forall r \geq 3\\
\end{align*}
With $x = \frac{w(H)}{n}$, the resulting approximation ratio is:
\[\frac{4 + (r - 2)(r - x - 1)}{2(r - x)}\]

\begin{figure}[t]
  \centering
  \includegraphics[width=12cm]{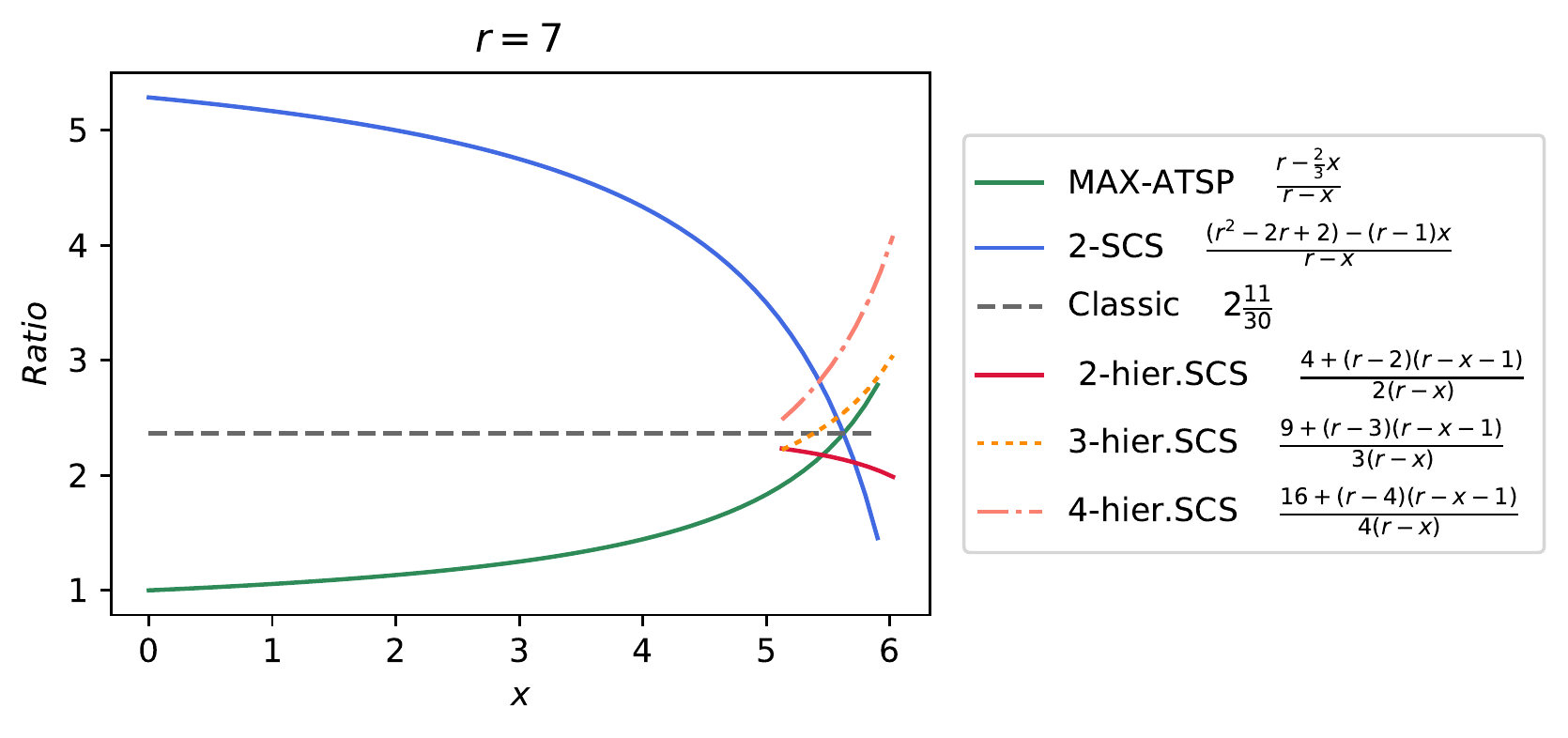}\\
  \includegraphics[width=12cm]{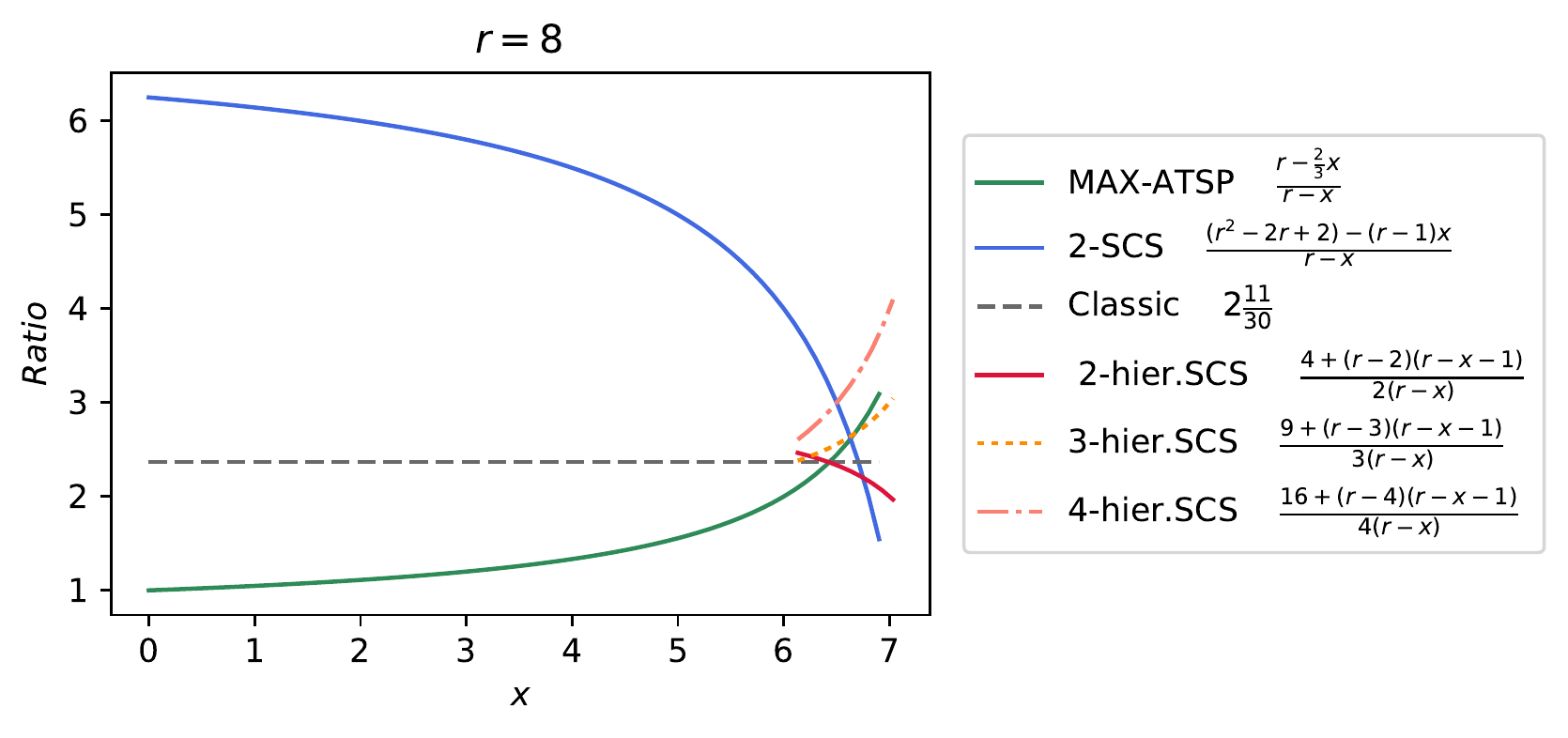}
  \caption{Extension of our approach by considering two additional levels, {\em i.e.} $(r-3)$ and $(r-4)$ overlaps.}
  
  \label{fig:extension}
\end{figure}

\noindent
We compare our bound to that of Golonev {\em et al.}, whose global ratio is of
$$ \alpha(r) = \max_{0 \leq x \leq r-1} \left \lbrace 
\mbox{min} \left \lbrace \frac{(r^2-2r+2)-(r-1)x}{r-x},\frac{r-\frac{2}{3}x)}{r-x} 
\right \rbrace
\right \rbrace \;.$$

\noindent
Our ratio $\beta(r)$ is of
$$\beta(r) = \max_{0 \leq x \leq r-1} \left \lbrace 
\mbox{min} \left \lbrace \frac{4 + (r - 2)(r - x - 1)}{2(r - x)},\frac{(r^2-2r+2)-(r-1)x}{r-x},\frac{r-\frac{2}{3}x}{r-x} 
\right \rbrace
\right \rbrace \;.$$

In Figure~\ref{fig:plots_r6_r9} we compare the MAX-ATSP bound, the best general
bound of $2\frac{11}{30}$, Golonev {et al.} bound and ours, for
$6 \leq r \leq 9$ and $ 0 \leq x \leq r.$ The plots show that our approach gives better results than Golovnev {\em et al.} for $5 < r
< 8$. However, beyond this limit of $r = 8$, our bound is not better than the $2 \frac{11}{30}$ general approximation ratio.

\section{Generalization of the hierarchical approach on more than two steps}

In order to push further our approach, we could consider three levels instead
of two, by taking into account $(r-3)$ edges in addition to $(r-1)$ and $(r-2)$ ones. Indeed, Property~\ref{prop:missingedge} can be directly extended to this case. However, our
method requires to bound the weight of all edges in our superstring
with the minimum overlap, $(r-3)$ in this case. This can be 
pushed even further by considering four levels, and so on. However, in
Figure~\ref{fig:extension} one can see that by extending the hierarchical approach, the approximation ratio becomes worse than that of the $2$-level approximation algorithm (presented in the previous section) for $r=7$ and $r=8$. Indeed, the approximation we introduced in the weight computation is too loose compared to the precision we gain by considering the additional levels.

{\small \bibliographystyle{plain}

}

\end{document}